\documentclass[12pt]{article}
\usepackage{amssymb,slashed,latexsym,amsmath,graphicx}
\newsavebox{\uuunit}
\sbox{\uuunit}
    {\setlength{\unitlength}{0.825em}
     \begin{picture}(0.6,0.7)
        \thinlines
        \put(0,0){\line(1,0){0.5}}
        \put(0.15,0){\line(0,1){0.7}}
        \put(0.35,0){\line(0,1){0.8}}
       \multiput(0.3,0.8)(-0.04,-0.02){12}{\rule{0.5pt}{0.5pt}}
     \end {picture}}

\csname @addtoreset\endcsname{equation}{section}

\long\def\symbolfootnote[#1]#2{\begingroup%
\def\thefootnote{\fnsymbol{footnote}}\footnote[#1]{#2}\endgroup}

\begin{document}

\begin{titlepage}
\begin{flushright}
CERN-PH-TH/2011-201\\
August 15, 2011\\
\end{flushright}
\vspace{.5cm}
\begin{center}
\baselineskip=16pt {\LARGE Higgs Decay into Two Photons, Revisited}\\
\vfill
{\large R.~Gastmans$^{1,}$\symbolfootnote[1]{Work
supported in part by the FWO-Vlaanderen, project G.0651.11, and in
part by the Federal Office for Scientific, Technical and Cultural
Affairs through the `Interuniversity Attraction Poles Programme --
Belgian Science Policy' P6/11-P.}}, {\large Sau Lan
Wu$^{2,}$\symbolfootnote[2]{Work supported in part
by the United States Department of Energy Grant No.~DE-FG02-95ER40896.}},
and {\large Tai Tsun Wu$^3$, 
  } \\

\vfill
{\small $^1$ Instituut voor Theoretische Fysica, Katholieke Universiteit Leuven,\\
       Celestijnenlaan 200D, B-3001 Leuven, Belgium. \\  \vspace{6pt}
$^2$ Department of Physics, University of Wisconsin, Madison WI 53706, USA.\\
\vspace{6pt} $^3$ Gordon McKay Laboratory, Harvard University,
Cambridge MA 02138, USA,\\ \vspace{6pt}
       and\\ \vspace{6pt}
Theory Division, CERN, CH-1211 Geneva 23, Switzerland.   \\[2mm] }
\end{center}
\vfill
\begin{center}
{\bf Abstract}
\end{center}
{\small The one-loop calculation of the amplitude for the Higgs
decay $H\rightarrow\gamma\gamma$ due to virtual $W$'s in the unitary
gauge is presented. As the Higgs does not directly couple to the
massless photons, the one-loop amplitude is finite. The calculation
is performed in a straightforward way, without encountering
divergences. In particular, artifacts like dimensional
regularization are avoided. This is achieved by judiciously routing
the external momenta through the loop and by combining the
integrands of the amplitudes before carrying out the integration
over the loop momentum. The present result satisfies the decoupling
theorem for infinite Higgs mass, and is thus different from the
earlier results obtained in the $\xi=1$ gauge using dimensional
regularization. The difference between the results is traced to the
use of dimensional regularization.}

\end{titlepage}
\addtocounter{page}{1}
\newpage
\section{Introduction}\label{sec1}
The Higgs decay into two photons,
\begin{equation}
H\rightarrow\gamma\gamma\,,\label{eq1.1}
\end{equation}
is an important channel in the search for the Higgs
particle~\cite{R1} at the Large Hadron Collider. If the Higgs mass
is near about~115~GeV$/c^2$, as favored by the first possible
evidence~\cite{R1a,R1b} from the Large Electron-Positron collider
(LEP) at CERN, then this decay is a good way to look for the Higgs,
because the photons can be seen cleanly.

The decay~\eqref{eq1.1} was studied theoretically many years
ago~\cite{R2,R3}. In particular, in the standard model of Glashow,
Weinberg, and Salam~\cite{R4}, there are two major contributions,
one from the top loop and one from the $W$ loop. Interestingly,
according to the previous calculations~\cite{R3}, these two
contributions are qualitatively different: while the one from the
top loop satisfies the decoupling theorem~\cite{R7}, that from
the~$W$ does not. In this context, decoupling is the phenomenon of a
particle to cease interaction with other particles when its mass
grows arbitrarily large.

Although there is no solid argument for the decoupling theorem, it
appears to be physically quite reasonable. With this in mind, it is
the purpose of this paper to revisit the one-loop~$W$ contribution
to the decay~\eqref{eq1.1}. To this end, we shall perform the
calculation in a way that differs substantially from the earlier
one, and we shall present the details in what follows.

The earliest calculation for the decay width is given by Ellis,
Gaillard, and Nanopoulos~\cite{R2}. That calculation can be
characterized as follows:
\renewcommand{\labelenumi}{(\alph{enumi})}
\begin{enumerate}
\item it is carried out in the~$R_\xi$ gauge with the
choice~$\xi=1$, \mbox{i.e.}, the~$R_1$ gauge;
\item dimensional regularization~\cite{R9} is used; and
\item the mass of the Higgs particle is taken to be much smaller than the~$W$ mass.
\end{enumerate}
Concerning~(a), in principle, all values of~$\xi$ are equivalent,
but the algebra is vastly simpler when the value~$\xi=1$ is chosen.
The~(b) is to be discussed extensively in this paper. While the
Higgs particle was believed to be much lighter than that of the~$W$
at the time this Ref.~\cite{R2} was written, it is now known this is
not so~\cite{R1a,R1b}.

The result obtained in Ref.~\cite{R2} has been confirmed by later
calculations~\cite{R3} and extended to arbitrary values of the Higgs
mass. Again, these are one-loop calculations in the $R_1$ gauge
using dimensional regularization --- see~(a) and~(b) above.

In contrast, the following different point of view is taken for the
present study. Since the photon is massless, there is no coupling of
the Higgs particle to the photon in the Lagrangian of the standard
model. Since the standard model is certainly one-loop
renormalizable~\cite{R10}, this absence of direct coupling implies
that the one-loop contribution to the decay~\eqref{eq1.1} through
the~$W$ loop must necessarily be finite. It is thus emphasized that
the quantity being studied is a finite one, and therefore it must be
possible to carry out the calculation in a completely
straightforward manner, without the introduction of, for example,
regularization of any kind. Accordingly, the~(a), (b), and (c) above
are to be replaced, for the present study, by the following:
\renewcommand{\labelenumi}{(\alph{enumi}')}
\begin{enumerate}
\item it is to be carried out in the simplest gauge;
\item throughout this study,
\begin{equation}
\mbox{the space-time dimension} = 4\,;\label{eq1.2a}
\end{equation}
and
\item the mass of the Higgs particle is arbitrary.
\end{enumerate}

The importance of the present study is due to the fact that the
result of the calculation on the basis of~(a'), (b'), and~(c') is
different from the previous one of~(a), (b), and~(c). Furthermore,
the present result does satisfy the decoupling theorem discussed
above.

In Sec.~\ref{sec2}, we present the formulae for the Feynman diagrams
in the unitary gauge; these are the ones we want to calculate. In
Sec.~\ref{sec3}, these amplitudes are simplified and combined to
give the present result that satisfies the decoupling theorem. Since
our result is different from the previous, generally accepted one,
we have chosen to present the calculation in detail, both for
eventual verification and because of  the unfamiliar nature of the
unitary gauge. Finally, in Sec.~\ref{sec4}, we trace back the reason
why our results is different from the previous one, and find that
the difference is due to~(b) versus~(b'). In other words, it is the
use of dimensional regularization that is the cause of the violation
of the decoupling theorem.

\section{Formulation of the problem}\label{sec2}

In this section, the problem is to be formulated in accordance with
the~(a'), (b'), and~(c') above. As to the choice of the simplest
gauge, a natural one is the unitary gauge --- this is the gauge
without any ghost and hence only the physical particles enter into
the perturbative calculation.  Nevertheless, this choice requires a
careful discussion because it is the conventional wisdom that the
unitary gauge is not suited for such calculations in quantum field
theory.

In the unitary gauge, the~$W$ propagator takes the form
\begin{equation}
P^{\alpha\beta}(p)=-i\,\frac{g^{\alpha\beta}-p^\alpha
p^\beta/M^2}{p^2-M^2+i\epsilon}\,,\label{eq1.2}
\end{equation}
where the quantity~$M$ is the mass of the vector particle~$W$ and
$p$ its four-momentum. Thus, this~$W$ propagator in the unitary
gauge consists of two terms: a first one that behaves
as~$(p^2)^{-1}$ for large~$p^2$, and a second one that behaves
as~$(p^2)^0$. The presence of this second term is the reason why the
unitary gauge is rarely used, due to the difficulties, in general,
of carrying out renormalization.

For the present problem, however, this difficulty does not enter,
because there is no divergence as discussed above, and hence there
is no need to renormalize. On the other hand, the absence of ghosts
in the unitary gauge greatly simplifies the necessary calculation.
There are only four relevant Feynman rules as given in
Fig.~\ref{fig1}, and these four Feynman rules lead to only three
\begin{figure}[t!]
\refstepcounter{figure} \label{fig1} \addtocounter{figure}{-1}
\begin{center}
\includegraphics[width=1.\textwidth]{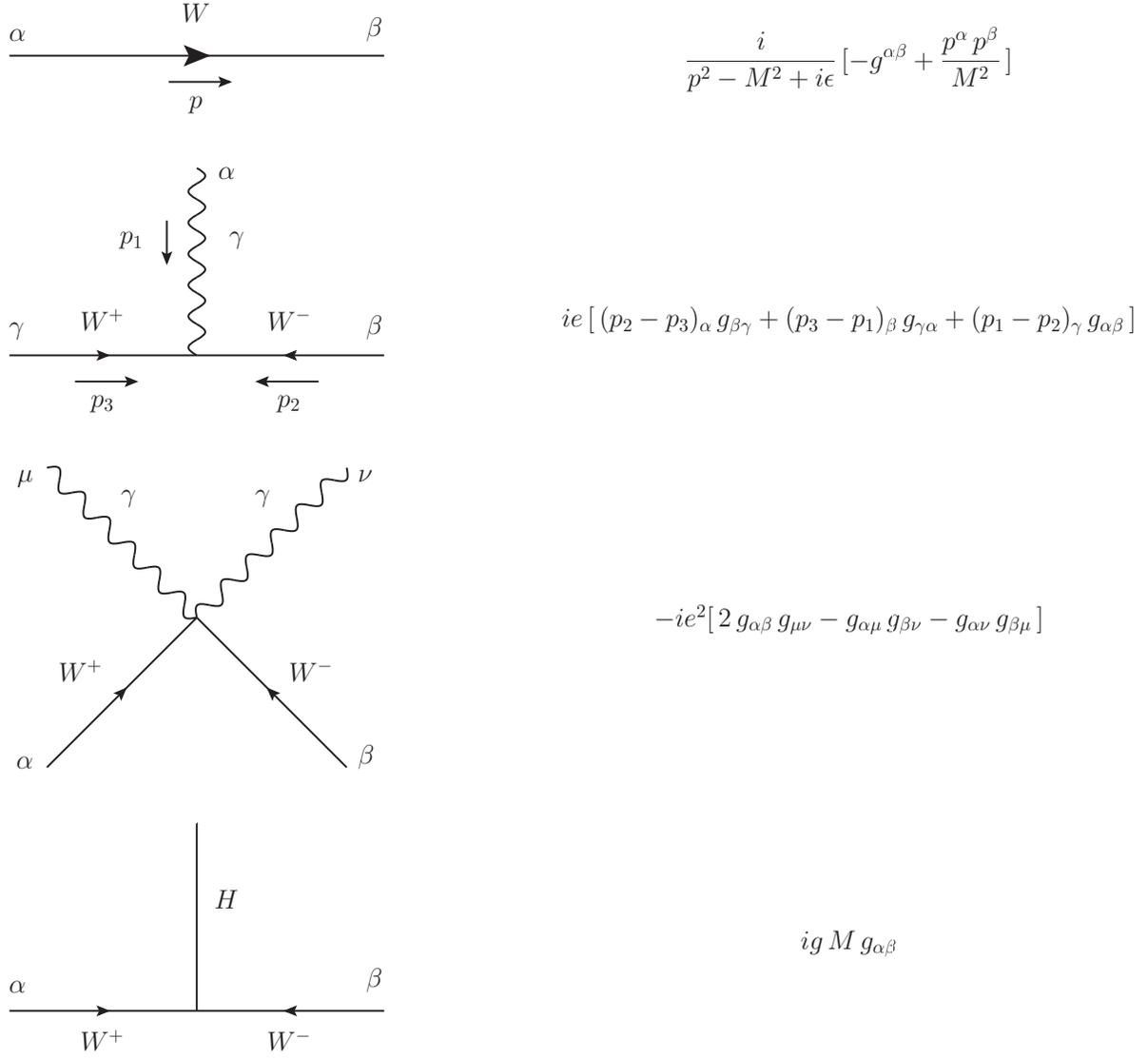}
\caption{The relevant Feynman rules in the unitary gauge for the
decay~$H\rightarrow\gamma\gamma$ at the one~$W$-loop level.}
\end{center}
\end{figure}
one~$W$-loop diagrams for the decay~\eqref{eq1.1} through a~$W$
loop. This is to be compared with fourteen Feynman diagrams in
the~$R_1$ gauge.

The three diagrams in the unitary gauge are shown in
Fig.~\ref{fig2}. This reduction of the number of Feynman diagrams to
be calculated is not free: the price to be paid is that, due to the
second term in the~$W$ propagator~\eqref{eq1.2}, these three
diagrams lead to integrals that are highly divergent. Because of
this divergence, the integrands for these three diagrams must be
added together before integrating with respect to the loop momentum.
Since shifting the momentum variable is not allowed for such
divergent integrals, the choices of the momentum variables for the
three diagrams are interdependent.

Fortunately, such interdependence of momentum choices between
different diagrams is well known in quantum field
theory~\cite{R12,R13,R14}, and has been studied for the present
problem~\cite{R15}, leading to the choice of momenta already shown
in Fig.~\ref{fig2}. It is now straightforward to write down the
amplitudes corresponding to these three diagrams.

The corresponding amplitudes are
\begin{eqnarray}
{\cal M}_1&=&\frac{-ie^2gM}{(2\pi)^4}\int d^4k\,[\,g_{\alpha}^{\beta}-(k+\dfrac{k_1+k_2}{2})_\alpha\,(k+\dfrac{k_1+k_2}{2})^\beta/M^2\,]\nonumber\\
&&\nonumber\\
&&\times[\,g^{\rho\sigma}-(k+\dfrac{-k_1+k_2}{2})^\rho\,(k+\dfrac{-k_1+k_2}{2})^\sigma/M^2\,]\nonumber\\
&&\nonumber\\
&&\times[\,g^{\alpha\gamma}-(k-\dfrac{k_1+k_2}{2})^\alpha\,(k-\dfrac{k_1+k_2}{2})^\gamma/M^2\,]\nonumber\\
&&\nonumber\\
&&\times[\,(k+\dfrac{3k_1+k_2}{2})_\rho\, g_{\beta\mu}+(k+\dfrac{-3k_1+k_2}{2})_\beta\, g_{\mu\rho}+(-2k-k_2)_\mu\, g_{\rho\beta}\,]\nonumber\\
&&\nonumber\\
&&\times\frac{(k-\dfrac{k_1+3k_2}{2})_\sigma\,g_{\gamma\nu}+(k+\dfrac{-k_1+3k_2}{2})_\gamma\,
g_{\nu\sigma}+(-2k+k_1)_\nu\, g_{\sigma\gamma}}
{\big[\big(k+\frac{k_1+k_2}{2}\big)^2-M^2+i\epsilon\big]\,\big[\big(k+\frac{-k_1+k_2}{2}\big)^2-M^2+i\epsilon\big]\,
\big[\big(k-\frac{k_1+k_2}{2}\big)^2-M^2+i\epsilon\big]}\,,\nonumber\\
&&\label{eq2.1}
\end{eqnarray}
\begin{figure}[t!]
\refstepcounter{figure} \label{fig2} \addtocounter{figure}{-1}
\begin{center}
\includegraphics[width=1.\textwidth]{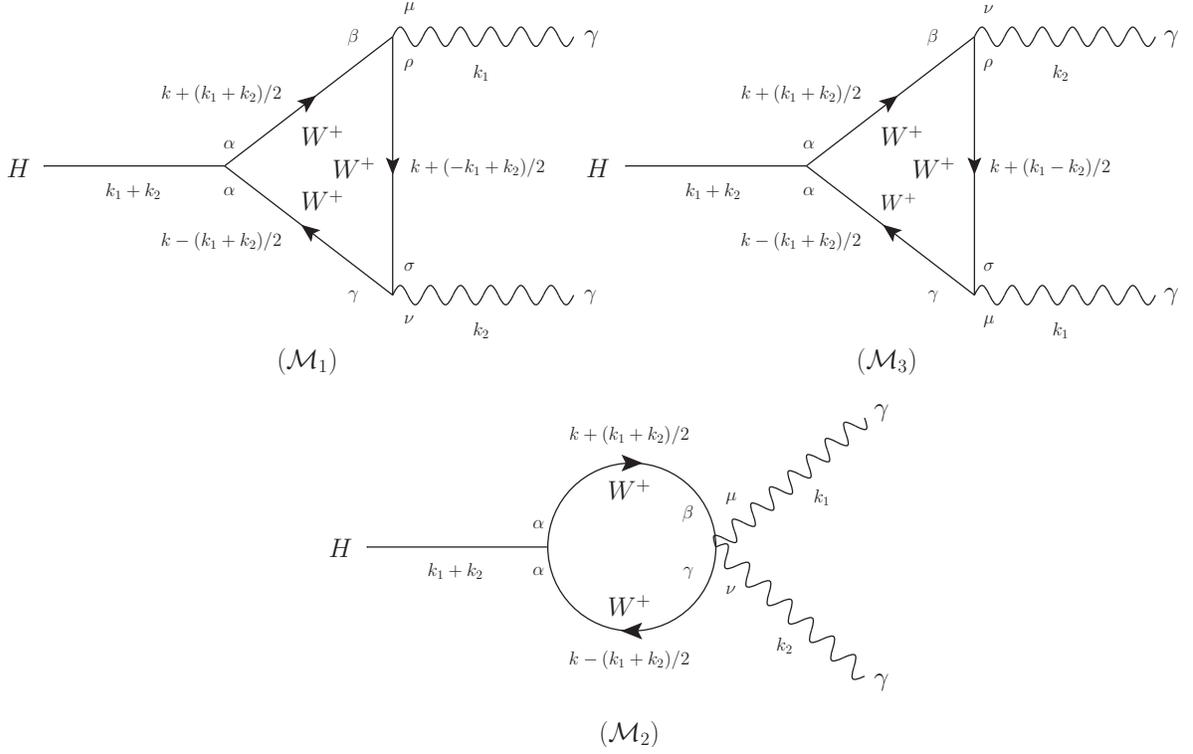}
\caption{The one-loop diagrams with virtual $W$'s in the unitary
gauge that contribute to the amplitude for
$H\rightarrow\gamma\gamma$.}
\end{center}
\end{figure}
\begin{eqnarray}
{\cal M}_2&=&\frac{ie^2gM}{(2\pi)^4}\int d^4k\,[\,g_{\alpha}^{\beta}-(k+\dfrac{k_1+k_2}{2})_\alpha\,(k+\dfrac{k_1+k_2}{2})^\beta/M^2\,]\nonumber\\
&&\nonumber\\
&&\times[\,g^{\alpha\gamma}-(k-\dfrac{k_1+k_2}{2})^\alpha\,(k-\dfrac{k_1+k_2}{2})^\gamma/M^2\,]\nonumber\\
&&\nonumber\\
&&\times\frac{2\,g_{\mu\nu}\,g_{\beta\gamma}-g_{\mu\beta}\,g_{\nu\gamma}-g_{\mu\gamma}\,g_{\nu\beta}}
{\big[\big(k+\frac{k_1+k_2}{2}\big)^2-M^2+i\epsilon\big]\,
\big[\big(k-\frac{k_1+k_2}{2}\big)^2-M^2+i\epsilon\big]}\,,
\label{eq2.2}
\end{eqnarray}
and
\begin{eqnarray}
{\cal M}_3&=&\frac{-ie^2gM}{(2\pi)^4}\int d^4k\,[\,g_{\alpha}^{\beta}-(k+\dfrac{k_1+k_2}{2})_\alpha\,(k+\dfrac{k_1+k_2}{2})^\beta/M^2\,]\nonumber\\
&&\nonumber\\
&&\times[\,g^{\rho\sigma}-(k+\dfrac{k_1-k_2}{2})^\rho\,(k+\dfrac{k_1-k_2}{2})^\sigma/M^2\,]\nonumber\\
&&\nonumber\\
&&\times[\,g^{\alpha\gamma}-(k-\dfrac{k_1+k_2}{2})^\alpha\,(k-\dfrac{k_1+k_2}{2})^\gamma/M^2\,]\nonumber\\
&&\nonumber\\
&&\times[\,(k+\dfrac{k_1+3k_2}{2})_\rho\, g_{\beta\nu}+(k+\dfrac{k_1-3k_2}{2})_\beta\, g_{\nu\rho}+(-2k-k_1)_\nu\, g_{\rho\beta}\,]\nonumber\\
&&\nonumber\\
&&\times\frac{(k-\dfrac{3k_1+k_2}{2})_\sigma\,g_{\gamma\mu}+(k+\dfrac{3k_1-k_2}{2})_\gamma\,
g_{\mu\sigma}+(-2k+k_2)_\mu\, g_{\sigma\gamma}}
{\big[\big(k+\frac{k_1+k_2}{2}\big)^2-M^2+i\epsilon\big]\,\big[\big(k+\frac{k_1-k_2}{2}\big)^2-M^2+i\epsilon\big]\,
\big[\big(k-\frac{k_1+k_2}{2}\big)^2-M^2+i\epsilon\big]}\,.\nonumber\\
&&\label{eq2.3}
\end{eqnarray}

In these formulae, $e$ is the electric charge and $g$ is the SU(2)
electroweak coupling constant. One should note that, in the
eqs.~\eqref{eq2.1}, \eqref{eq2.2}, and \eqref{eq2.3}, we have
omitted the polarization vectors $(\epsilon^\mu)^*$ and
$(\epsilon^\nu)^*$ for the outgoing photons. Also, since we are
dealing with real photons, we have
\begin{equation}
k^2_1=k^2_2=0\,,\hspace{3cm}k_{1\mu}=k_{2\nu}=0\,,\label{eq2.4}
\end{equation}
where $k_1$ and $k_2$ are the four-momenta of the two photons, and
$k_1+k_2$ is the four-momentum of the Higgs particle. Consequently,
\begin{equation}
2\,(k_1\cdot k_2)=M_H^2\,,\label{eq2.5}
\end{equation}
where $M_H$ is the Higgs mass. Also note that the amplitude ${\cal M}_2$ is
symmetrical for the interchange of the two photons 1 and 2
(\,$k_1\leftrightarrow k_2\,,\, \mu\leftrightarrow\nu\,$).

For the evaluation of the amplitude, use is made of the Ward identities to simplify the algebra. For the~$WW\gamma$ vertex
\begin{equation}
V_{\alpha\beta\gamma}(p_1,p_2,p_3)=(p_2-p_3)_\alpha\, g_{\beta\gamma}+(p_3-p_1)_\beta\, g_{\gamma\alpha}
+(p_1-p_2)_\gamma\, g_{\alpha\beta}\,,\label{eq2.6}
\end{equation}
with all four-momenta $p_1, p_2, p_3\ (p_1+p_2+p_3=0)$ incoming,
this identity reads
\begin{equation}
p_1^\alpha\, V_{\alpha\beta\gamma}(p_1,p_2,p_3)=[\,p_3^2\, g_{\beta\gamma}-p_{3\beta}p_{3\gamma}\,]-
[\,p_2^2\, g_{\beta\gamma}-p_{2\beta}p_{2\gamma}\,]\,.\label{eq2.7}
\end{equation}
For the special case that $p_2$, \mbox{e.g.}, is associated with one
of the real outgoing photons, this identity reduces to
\begin{equation}
p_1^\alpha\, V_{\alpha\mu\gamma}(p_1,-k_1,p_3)=p_3^2\,
g_{\mu\gamma}-p_{3\mu}\,p_{3\gamma}\,,\label{eq2.8}
\end{equation}
because of the relations~\eqref{eq2.4}. Of course, there is a
similar relation for photon~2. In practice, we often use a slightly
modified version of this equation in Sec.~\ref{sec3}, \mbox{i.e.},
\begin{equation}
p_1^\alpha\, V_{\alpha\mu\gamma}(p_1,-k_1,p_3)=[\,p_3^2-M^2\,]\,
g_{\mu\gamma}-p_{3\mu}\,p_{3\gamma}+M^2\,g_{\mu\gamma}\,,\label{eq2.9}
\end{equation}
the reason being that the first term in~\eqref{eq2.9} can be used to
cancel a factor in the denominator of ${\cal M}_1$ or ${\cal M}_2$. In this way,
cancelations with contributions from ${\cal M}_2$ can be achieved.

Finally, from~\eqref{eq2.9}, it immediately follows that
\begin{equation}
p_1^\alpha\,p_3^\gamma\,
V_{\alpha\mu\gamma}(p_1,-k_1,p_3)=0\,,\label{eq2.10}
\end{equation}
and, similarly,
\begin{equation}
p_1^\alpha\,p_3^\gamma\,
V_{\alpha\nu\gamma}(p_1,-k_2,p_3)=0\,.\label{eq2.11}
\end{equation}

\section{The evaluation of the amplitude}\label{sec3}
Our procedure for the evaluation of the amplitude is
straightforward, but somewhat lengthy. We examine successively the
terms in~$M^{-n}\,,\ n=6,4,2,0$ in~${\cal M}_1$, ${\cal M}_2$,
and~${\cal M}_3$ using the Ward identities listed in
Sec.~\ref{sec2}. Here, a term in~$M^{-n}$ means a term with an
explicit overall factor of~$M^{-n}$, not counting the~$M$ in the
factor~$\pm ie^2gM$ in eqs.~\eqref{eq2.1}-\eqref{eq2.3}. We shall
find that all the terms with negative powers of~$M$ give a vanishing
contribution. The resulting amplitude is then seen to satisfy the
decoupling theorem~\cite{R7}.
\subsection{The terms in $M^{-6}$}\label{sec3.1}
In ${\cal M}_1$ as given by~\eqref{eq2.1}, there is only one term
proportional to $M^{-6}$. It is obtained by taking the longitudinal
parts of all three $W$-propagators. It then follows that its
contribution vanishes because of the Ward identities~\eqref{eq2.10}
or~\eqref{eq2.11}. A similar conclusion holds for ${\cal M}_3$ given
by~\eqref{eq2.3}. As~${\cal M}_2$ from eq.~\eqref{eq2.2} has no terms in
$M^{-6}$, it follows that the entire amplitude has no such terms.
\subsection{The terms in $M^{-4}$}\label{sec3.2}
The terms in ${\cal M}_1$ proportional to $M^{-4}$ necessarily
result from the combination of two longitudinal parts of
propagators. Because of the Ward identities~\eqref{eq2.10}
and~\eqref{eq2.11}, only the two propagators adjacent to the Higgs
vertex contribute. They give
\begin{eqnarray}
{\cal M}_{11}&=&\dfrac{-ie^2gM}{(2\pi)^4}\,\dfrac{1}{M^4}\int
d^4k\,(k+\dfrac{k_1+k_2}{2})_\alpha\,(k+\dfrac{k_1+k_2}{2})^\beta\,
(k-\frac{k_1+k_2}{2})^\alpha\,(k-\frac{k_1+k_2}{2})^\gamma\nonumber\\
&&\nonumber\\
&&\times\,g^{\rho\sigma}\,[\,(k+\dfrac{3k_1+k_2}{2})_\rho\, g_{\beta\mu}+
(k+\dfrac{-3k_1+k_2}{2})_\beta\, g_{\mu\rho}+(-2k-k_2)_\mu\, g_{\rho\beta}\,]\nonumber\\
&&\nonumber\\
&&\times\frac{(k-\dfrac{k_1+3k_2}{2})_\sigma\,g_{\gamma\nu}+(k+\dfrac{-k_1+3k_2}{2})_\gamma\,
g_{\nu\sigma}+(-2k+k_1)_\nu\,g_{\sigma\gamma}}
{\big[\big(k+\frac{k_1+k_2}{2}\big)^2-M^2+i\epsilon\big]\,\big[\big(k+\frac{-k_1+k_2}{2}\big)^2-M^2+i\epsilon\big]\,
\big[\big(k-\frac{k_1+k_2}{2}\big)^2-M^2+i\epsilon\big]}\,,\nonumber\\
&&\label{eq3.1}
\end{eqnarray}
Using the Ward identity~\eqref{eq2.9}, this expression can be
rewritten in three terms
\begin{equation}
{\cal M}_{11}={\cal M}_{111}+{\cal M}_{112}+{\cal
M}_{113},,\label{eq3.2}
\end{equation}
with
\begin{eqnarray}
{\cal M}_{111}&=&\dfrac{-ie^2gM}{(2\pi)^4}\,\dfrac{1}{M^4}\int
d^4k\,(k+\dfrac{k_1+k_2}{2})_\alpha\,(k-\dfrac{k_1+k_2}{2})^\alpha\,(k-\dfrac{k_1+k_2}{2})^\gamma\,g^{\rho\sigma}\,g_{\rho\mu}\nonumber\\
&&\nonumber\\
&&\times\frac{(k-\dfrac{k_1+3k_2}{2})_\sigma\,g_{\gamma\nu}+(k+\dfrac{-k_1+3k_2}{2})_\gamma\,
g_{\nu\sigma}+(-2k+k_1)_\nu\,g_{\sigma\gamma}}
{\big[\big(k+\frac{k_1+k_2}{2}\big)^2-M^2+i\epsilon\big]\,
\big[\big(k-\frac{k_1+k_2}{2}\big)^2-M^2+i\epsilon\big]}\nonumber\\
&&\nonumber\\
&&\nonumber\\
&=&\dfrac{-ie^2gM}{(2\pi)^4}\,\dfrac{1}{M^4}\int
d^4k\,[\,k^2-\frac{(k_1\cdot k_2)}{2}\,]\,(k-\frac{k_1+k_2}{2})^\gamma\nonumber\\
&&\nonumber\\
&&\times\frac{(k-\dfrac{3k_2}{2})_\mu\,g_{\gamma\nu}+(k+\dfrac{-k_1+3k_2}{2})_\gamma\,
g_{\mu\nu}+(-2k+k_1)_\nu\, g_{\mu\gamma}}
{\big[\big(k+\frac{k_1+k_2}{2}\big)^2-M^2+i\epsilon\big]\,
\big[\big(k-\frac{k_1+k_2}{2}\big)^2-M^2+i\epsilon\big]}\,,\label{eq3.2a}
\end{eqnarray}
\begin{eqnarray}
{\cal M}_{112}&=&\dfrac{ie^2gM}{(2\pi)^4}\,\dfrac{1}{M^4}\int
d^4k\,[\,k^2-\dfrac{(k_1\cdot
k_2)}{2}\,]\,(k-\dfrac{k_1+k_2}{2})^\gamma\,(k+\dfrac{-k_1+k_2}{2})^\sigma\,
(k+\dfrac{k_2}{2})_\mu\nonumber\\
&&\nonumber\\
&&\times\frac{(k-\dfrac{k_1+3k_2}{2})_\sigma\,g_{\gamma\nu}+(k+\dfrac{-k_1+3k_2}{2})_\gamma\,
g_{\nu\sigma}+(-2k+k_1)_\nu\, g_{\sigma\gamma}}
{\big[\big(k+\frac{k_1+k_2}{2}\big)^2-M^2+i\epsilon\big]\,\big[\big(k+\frac{-k_1+k_2}{2}\big)^2-M^2+i\epsilon\big]\,
\big[\big(k-\frac{k_1+k_2}{2}\big)^2-M^2+i\epsilon\big]}\,,\nonumber\\
&&\label{eq3.2b}
\end{eqnarray}
and
\begin{eqnarray}
{\cal M}_{113}&=&\dfrac{-ie^2gM}{(2\pi)^4}\,\dfrac{1}{M^2}\int
d^4k\,[\,k^2-\dfrac{(k_1\cdot k_2)}{2}\,]\,(k-\dfrac{k_1+k_2}{2})^\gamma\nonumber\\
&&\nonumber\\
&&\times\frac{(k-\dfrac{3k_2}{2})_\mu\,g_{\gamma\nu}+(k+\dfrac{-k_1+3k_2}{2})_\gamma\,
g_{\mu\nu}+(-2k+k_1)_\nu\, g_{\mu\gamma}}
{\big[\big(k+\frac{k_1+k_2}{2}\big)^2-M^2+i\epsilon\big]\,\big[\big(k+\frac{-k_1+k_2}{2}\big)^2-M^2+i\epsilon\big]\,
\big[\big(k-\frac{k_1+k_2}{2}\big)^2-M^2+i\epsilon\big]}\,.\nonumber\\
&&\label{eq3.3}
\end{eqnarray}
The last contribution, ${\cal M}_{113}$, will be treated in
subsection~\ref{sec3.3} together with the other terms in $M^{-2}$.

First, we apply the Ward identity~\eqref{eq2.8} to ${\cal M}_{111}$:
\begin{equation}
{\cal M}_{111}={\cal M}_{1111}+{\cal M}_{1112}\,,\label{eq3.3a}
\end{equation}
with
\begin{eqnarray}
{\cal M}_{1111}&=&\dfrac{-ie^2gM}{(2\pi)^4}\,\dfrac{1}{M^4}\int
d^4k\,\frac{[\,k^2-\dfrac{(k_1\cdot
k_2)}{2}\,]\,g_{\mu\nu}\,\big(k+\dfrac{-k_1+k_2}{2}\big)^2}
{\big[\big(k+\frac{k_1+k_2}{2}\big)^2-M^2+i\epsilon\big]\,
\big[\big(k-\frac{k_1+k_2}{2}\big)^2-M^2+i\epsilon\big]}\nonumber\\
&&\nonumber\\
&=&\dfrac{-ie^2gM}{(2\pi)^4}\,\dfrac{1}{M^4}\int
d^4k\,\frac{[\,k^2-\dfrac{(k_1\cdot k_2)}{2}\,]\,g_{\mu\nu}\,
[\,k^2-(k\cdot k_1)+(k\cdot k_2)-\dfrac{(k_1\cdot k_2)}{2}\,]}
{\big[\big(k+\frac{k_1+k_2}{2}\big)^2-M^2+i\epsilon\big]\,
\big[\big(k-\frac{k_1+k_2}{2}\big)^2-M^2+i\epsilon\big]}\nonumber\\
&&\label{eq3.3b}
\end{eqnarray}
and
\begin{eqnarray}
{\cal M}_{1112}&=&\dfrac{ie^2gM}{(2\pi)^4}\,\dfrac{1}{M^4}\int
d^4k\,\frac{[\,k^2-\dfrac{(k_1\cdot
k_2)}{2}\,]\,(k+\dfrac{k_2}{2})_\mu\,(k-\dfrac{k_1}{2})_\nu}
{\big[\big(k+\frac{k_1+k_2}{2}\big)^2-M^2+i\epsilon\big]\,
\big[\big(k-\frac{k_1+k_2}{2}\big)^2-M^2+i\epsilon\big]}\,.\label{eq3.3c}
\end{eqnarray}
From the amplitude ${\cal M}_3$ [see Fig.~\ref{fig2}], we obtain the
analogous expressions ${\cal M}_{3111}$ and ${\cal M}_{3112}$ by the
interchange of photons 1 and 2, \mbox{i.e.}, $k_1\leftrightarrow
k_2$ and $\mu\leftrightarrow\nu$. Hence,
\begin{equation}
{\cal M}_{1111}+{\cal
M}_{3111}=\dfrac{-ie^2gM}{(2\pi)^4}\,\dfrac{2}{M^4}\int
d^4k\,\frac{[\,k^2-\dfrac{(k_1\cdot
k_2)}{2}\,]\,g_{\mu\nu}\,[\,k^2-\dfrac{(k_1\cdot k_2)}{2}\,]}
{\big[\big(k+\frac{k_1+k_2}{2}\big)^2-M^2+i\epsilon\big]\,
\big[\big(k-\frac{k_1+k_2}{2}\big)^2-M^2+i\epsilon\big]}\label{eq3.3d}
\end{equation}
and
\begin{equation}
{\cal M}_{1112}+{\cal
M}_{3112}=\dfrac{ie^2gM}{(2\pi)^4}\,\dfrac{2}{M^4}\int
d^4k\,\frac{[\,k^2-\dfrac{(k_1\cdot
k_2)}{2}\,]\,(k_\mu\,k_\nu-\dfrac{k_{2\mu}\,k_{1\nu}}{4})}
{\big[\big(k+\frac{k_1+k_2}{2}\big)^2-M^2+i\epsilon\big]\,
\big[\big(k-\frac{k_1+k_2}{2}\big)^2-M^2+i\epsilon\big]}\,.\label{eq3.3e}
\end{equation}
However, the amplitude ${\cal M}_2$ also yields terms of order $M^{-4}$. They are
\begin{equation}
{\cal M}_{21}=\dfrac{ie^2gM}{(2\pi)^4}\,\dfrac{1}{M^4}\int
d^4k\,\frac{\big[\,k^2-\dfrac{(k_1\cdot
k_2)}{2}\,\big]\,\big[\,2g_{\mu\nu}\, \big(k^2-\dfrac{(k_1\cdot
k_2)}{2}\big)-2k_\mu\,k_\nu+\dfrac{k_{2\mu}\,k_{1\nu}}{2}\big]}
{\big[\big(k+\frac{k_1+k_2}{2}\big)^2-M^2+i\epsilon\big]\,
\big[\big(k-\frac{k_1+k_2}{2}\big)^2-M^2+i\epsilon\big]}\,.\label{eq3.3f}
\end{equation}
Combining the results from~\eqref{eq3.3d}, \eqref{eq3.3e},
and~\eqref{eq3.3f} then gives
\begin{equation}
{\cal M}_{1111}+{\cal M}_{3111}+{\cal M}_{1112}+{\cal
M}_{3112}+{\cal M}_{21}=0\,.\label{eq3.3g}
\end{equation}

Next, we again apply the Ward identity~\eqref{eq2.8}, this time to
the expression ${\cal M}_{112}$ as given by~\eqref{eq3.2b} yielding
two terms, \mbox{i.e.},
\begin{equation}
{\cal M}_{112}={\cal M}_{1121}+{\cal M}_{1122}\,,\label{eq3.4}
\end{equation}
with
\begin{eqnarray}
{\cal M}_{1121}&=&\dfrac{ie^2gM}{(2\pi)^4}\,\dfrac{1}{M^4}\int
d^4k\,{[\,k^2-\dfrac{(k_1\cdot
k_2)}{2}\,]\,(k+\dfrac{k_2}{2})_\mu\,(k-\dfrac{k_1}{2})_\nu}\nonumber\\
&&\nonumber\\
&&\times\frac{(k-\dfrac{k_1+k_2}{2})^2}
{\big[\big(k+\frac{k_1+k_2}{2}\big)^2-M^2+i\epsilon\big]\,\big[\big(k+\frac{-k_1+k_2}{2}\big)^2-M^2+i\epsilon\big]\,
\big[\big(k-\frac{k_1+k_2}{2}\big)^2-M^2+i\epsilon\big]}\nonumber\\
&&\label{eq3.5}
\end{eqnarray}
and
\begin{eqnarray}
{\cal M}_{1122}&=&\dfrac{-ie^2gM}{(2\pi)^4}\,\dfrac{1}{M^4}\int
d^4k\,[\,k^2-\dfrac{(k_1\cdot
k_2)}{2}\,]\,(k-\dfrac{k_1+k_2}{2})^\gamma\,(k+\dfrac{k_2}{2})_\mu\,
(k-\dfrac{k_1}{2})_\nu\nonumber\\
&&\nonumber\\
&&\times\frac{(k-\dfrac{k_1+k_2}{2})_\gamma}
{\big[\big(k+\frac{k_1+k_2}{2}\big)^2-M^2+i\epsilon\big]\,\big[\big(k+\frac{-k_1+k_2}{2}\big)^2-M^2+i\epsilon\big]\,
\big[\big(k-\frac{k_1+k_2}{2}\big)^2-M^2+i\epsilon\big]}\,.\nonumber\\
&&\label{eq3.6}
\end{eqnarray}
It is readily seen that the these two terms~\eqref{eq3.5}
and~\eqref{eq3.6} cancel:
\begin{equation}
{\cal M}_{1121}+{\cal M}_{1122}=0\,.\label{eq3.8}
\end{equation}
Hence, by virtue of eq.~\eqref{eq3.4},
\begin{equation}
{\cal M}_{112}=0\,.\label{eq3.9}
\end{equation}

We have thus shown that all the terms of order~$M^{-4}$ cancel. In
the process, we generated a term of order~$M^{-2}$
[see~\eqref{eq3.3}], which will have to be combined with the terms
to be treated in the next subsection~\ref{sec3.3}.

\subsection{The terms in $M^{-2}$}\label{sec3.3}
Because there are three $W$ propagators in the first Feynman diagram, we can distinguish
three contributions of order~$M^{-2}$ from the first amplitude~${\cal M}_1$. They are
\begin{eqnarray}
{\cal M}_{12}&=&\frac{ie^2gM}{(2\pi)^4}\,\frac{1}{M^2}\int
d^4k\,g^{\rho\sigma}\,(k-\frac{k_1+k_2}{2})^\beta\,(k-\frac{k_1+k_2}{2})^\gamma\nonumber\\
&&\nonumber\\
&&\times[\,(k+\frac{3k_1+k_2}{2})_\rho\, g_{\beta\mu}+(k+\frac{-3k_1+k_2}{2})_\beta\, g_{\mu\rho}+(-2k-k_2)_\mu\, g_{\rho\beta}\,]\nonumber\\
&&\nonumber\\
&&\times\frac{(k-{\displaystyle\frac{k_1+3k_2}{2})_\sigma\,g_{\gamma\nu}+(k+\frac{-k_1+3k_2}{2})_\gamma\,
g_{\nu\sigma}+(-2k+k_1)_\nu\,g_{\sigma\gamma}}}
{\big[\big(k+\frac{k_1+k_2}{2}\big)^2-M^2+i\epsilon\big]\,\big[\big(k+\frac{-k_1+k_2}{2}\big)^2-M^2+i\epsilon\big]\,
\big[\big(k-\frac{k_1+k_2}{2}\big)^2-M^2+i\epsilon\big]}\,,\nonumber\\
&&\label{eq4.1}
\end{eqnarray}
\begin{eqnarray}
{\cal M}_{13}&=&\frac{ie^2gM}{(2\pi)^4}\,\frac{1}{M^2}\int
d^4k\,g^{\beta\gamma}\,(k+\frac{-k_1+k_2}{2})^\rho\,(k+\frac{-k_1+k_2}{2})^\sigma\nonumber\\
&&\nonumber\\
&&\times[\,(k+\frac{3k_1+k_2}{2})_\rho\, g_{\beta\mu}+(k+\frac{-3k_1+k_2}{2})_\beta\, g_{\mu\rho}+(-2k-k_2)_\mu\, g_{\rho\beta}\,]\nonumber\\
&&\nonumber\\
&&\times\frac{(k-{\displaystyle\frac{k_1+3k_2}{2})_\sigma\,g_{\gamma\nu}+(k+\frac{-k_1+3k_2}{2})_\gamma\,
g_{\nu\sigma}+(-2k+k_1)_\nu\,g_{\sigma\gamma}}}
{\big[\big(k+\frac{k_1+k_2}{2}\big)^2-M^2+i\epsilon\big]\,\big[\big(k+\frac{-k_1+k_2}{2}\big)^2-M^2+i\epsilon\big]\,
\big[\big(k-\frac{k_1+k_2}{2}\big)^2-M^2+i\epsilon\big]}\,,\nonumber\\
&&\label{eq4.2}
\end{eqnarray}
and
\begin{eqnarray}
{\cal M}_{14}&=&\frac{ie^2gM}{(2\pi)^4}\,\frac{1}{M^2}\int
d^4k\,g^{\rho\sigma}\,(k+\frac{k_1+k_2}{2})^\gamma\,(k+\frac{k_1+k_2}{2})^\beta\nonumber\\
&&\nonumber\\
&&\times[\,(k+\frac{3k_1+k_2}{2})_\rho\, g_{\beta\mu}+(k+\frac{-3k_1+k_2}{2})_\beta\, g_{\mu\rho}+(-2k-k_2)_\mu\, g_{\rho\beta}\,]\nonumber\\
&&\nonumber\\
&&\times\frac{(k-{\displaystyle\frac{k_1+3k_2}{2})_\sigma\,g_{\gamma\nu}+(k+\frac{-k_1+3k_2}{2})_\gamma\,
g_{\nu\sigma}+(-2k+k_1)_\nu\,g_{\sigma\gamma}}}
{\big[\big(k+\frac{k_1+k_2}{2}\big)^2-M^2+i\epsilon\big]\,\big[\big(k+\frac{-k_1+k_2}{2}\big)^2-M^2+i\epsilon\big]\,
\big[\big(k-\frac{k_1+k_2}{2}\big)^2-M^2+i\epsilon\big]}\,,\nonumber\\
&&\label{eq4.3}
\end{eqnarray}
Applying the Ward identity~\eqref{eq2.9} to~${\cal M}_{12}$ yields
three terms, \mbox{i.e.},
\begin{equation}
{\cal M}_{12}={\cal M}_{121}+{\cal M}_{122}+{\cal
M}_{123}\,,\label{eq4.4}
\end{equation}
with
\begin{eqnarray}
{\cal M}_{121}&=&\frac{ie^2gM}{(2\pi)^4}\,\frac{1}{M^2}\int
d^4k\,g^\rho_\nu\,(k-\frac{k_1+k_2}{2})^\beta\nonumber\\
&&\nonumber\\
&&\times\frac{(k+{\displaystyle\frac{3k_1+k_2}{2})_\rho\,g_{\beta\mu}+
(k+\frac{-3k_1+k_2}{2})_\beta\,g_{\mu\rho}+(-2k-k_2)_\mu\,g_{\beta\rho}}}
{\big[\big(k+\frac{k_1+k_2}{2}\big)^2-M^2+i\epsilon\big]\,
\big[\big(k-\frac{k_1+k_2}{2}\big)^2-M^2+i\epsilon\big]}\,,\label{eq4.5}
\end{eqnarray}
\begin{eqnarray}
{\cal M}_{122}&=&\frac{-ie^2gM}{(2\pi)^4}\,\frac{1}{M^2}\int
d^4k\,(k-\frac{k_1+k_2}{2})^\beta\,(k+\frac{-k_1+k_2}{2})^\rho\,(k+\frac{-k_1+k_2}{2})_\nu\nonumber\\
&&\nonumber\\
&&\times\frac{(k+{\displaystyle\frac{3k_1+k_2}{2})_\rho\,g_{\beta\mu}+
(k+\frac{-3k_1+k_2}{2})_\beta\,g_{\mu\rho}+(-2k-k_2)_\mu\,g_{\beta\rho}}}
{\big[\big(k+\frac{k_1+k_2}{2}\big)^2-M^2+i\epsilon\big]\,[\big(k+\frac{-k_1+k_2}{2}\big)^2-M^2+i\epsilon\big]\,
\big[\big(k-\frac{k_1+k_2}{2}\big)^2-M^2+i\epsilon\big]}\,,\nonumber\\
&&\label{eq4.6}
\end{eqnarray}
and
\begin{eqnarray}
{\cal M}_{123}&=&\frac{ie^2gM}{(2\pi)^4}\,\int
d^4k\,(k-\frac{k_1+k_2}{2})^\beta\nonumber\\
&&\nonumber\\
&&\times\frac{(k+{\displaystyle\frac{3k_1+k_2}{2})_\nu\,g_{\beta\mu}+
(k+\frac{-3k_1+k_2}{2})_\beta\,g_{\mu\nu}+(-2k-k_2)_\mu\,g_{\beta\nu}}}
{\big[\big(k+\frac{k_1+k_2}{2}\big)^2-M^2+i\epsilon\big]\,[\big(k+\frac{-k_1+k_2}{2}\big)^2-M^2+i\epsilon\big]\,
\big[\big(k-\frac{k_1+k_2}{2}\big)^2-M^2+i\epsilon\big]}\,.\nonumber\\
&&\label{eq4.7}
\end{eqnarray}
Adding to ${\cal M}_{121}$ the analogous $1\leftrightarrow2$ term
from the amplitude~${\cal M}_3$ yields
\begin{eqnarray}
{\cal M}_{121}+{\cal
M}_{321}&=&\frac{ie^2gM}{(2\pi)^4}\,\frac{1}{M^2}\int
d^4k\,(k-\frac{k_1+k_2}{2})^\beta\nonumber\\
&&\nonumber\\
&&\times\frac{{\displaystyle(-k+\frac{k_1}{2})_\nu\,g_{\beta\mu}+
(2k-k_1-k_2)_\beta\,g_{\mu\nu}+(-k+\frac{k_2}{2})_\mu\,g_{\beta\nu}}}
{\big[\big(k+\frac{k_1+k_2}{2}\big)^2-M^2+i\epsilon\big]\,
\big[\big(k-\frac{k_1+k_2}{2}\big)^2-M^2+i\epsilon\big]}\,,\label{eq4.8}
\end{eqnarray}

However, also the amplitude ${\cal M}_2$ has terms of order $M^{-2}$:
\begin{eqnarray}
{\cal M}_{22}&=&\frac{-ie^2gM}{(2\pi)^4}\,\frac{1}{M^2}\int
d^4k\,(k+\frac{k_1+k_2}{2})^\gamma\,(k+\frac{k_1+k_2}{2})^\beta\nonumber\\
&&\nonumber\\
&&\times\frac{2\,g_{\mu\nu}\,g_{\beta\gamma}-g_{\mu\beta}\,g_{\nu\gamma}-g_{\mu\gamma}\,
g_{\nu\beta}}
{\big[\big(k+\frac{k_1+k_2}{2}\big)^2-M^2+i\epsilon\big]\,
\big[\big(k-\frac{k_1+k_2}{2}\big)^2-M^2+i\epsilon\big]}\,,\label{eq4.9}
\end{eqnarray}
and
\begin{eqnarray}
{\cal M}_{23}&=&\frac{-ie^2gM}{(2\pi)^4}\,\frac{1}{M^2}\int
d^4k\,(k-\frac{k_1+k_2}{2})^\beta\,(k-\frac{k_1+k_2}{2})^\gamma\nonumber\\
&&\nonumber\\
&&\times\frac{2\,g_{\mu\nu}\,g_{\beta\gamma}-g_{\mu\beta}\,g_{\nu\gamma}-g_{\mu\gamma}\,
g_{\nu\beta}}
{\big[\big(k+\frac{k_1+k_2}{2}\big)^2-M^2+i\epsilon\big]\,
\big[\big(k-\frac{k_1+k_2}{2}\big)^2-M^2+i\epsilon\big]}\,,\label{eq4.10}
\end{eqnarray}
From~\eqref{eq4.8} and~\eqref{eq4.10}, it is readily seen that
\begin{equation}
{\cal M}_{121}+{\cal M}_{321}+{\cal M}_{23}=0\,.\label{eq4.11}
\end{equation}

What we did for ${\cal M}_{12}$ and ${\cal M}_{23}$ can be repeated,
{\it mutatis mutandis}, for ${\cal M}_{14}$ and ${\cal M}_{22}$. The
Ward identity~\eqref{eq2.9} on ${\cal M}_{14}$ yields three terms
\begin{equation}
{\cal M}_{14}={\cal M}_{141}+{\cal M}_{142}+{\cal
M}_{143}\,,\label{eq4.12}
\end{equation}
with
\begin{eqnarray}
{\cal M}_{141}&=&\frac{ie^2gM}{(2\pi)^4}\,\frac{1}{M^2}\int
d^4k\,g^\sigma_\mu\,(k+\frac{k_1+k_2}{2})^\gamma\nonumber\\
&&\nonumber\\
&&\times\frac{(k-{\displaystyle\frac{k_1+3k_2}{2})_\sigma\,g_{\gamma\nu}+
(k+\frac{-k_1+3k_2}{2})_\gamma\,g_{\nu\sigma}+(-2k+k_1)_\nu\,g_{\gamma\sigma}}}
{\big[\big(k+\frac{k_1+k_2}{2}\big)^2-M^2+i\epsilon\big]\,
\big[\big(k-\frac{k_1+k_2}{2}\big)^2-M^2+i\epsilon\big]}\,,\label{eq4.13}
\end{eqnarray}
\begin{eqnarray}
{\cal M}_{142}&=&\frac{-ie^2gM}{(2\pi)^4}\,\frac{1}{M^2}\int
d^4k\,(k+\frac{k_1+k_2}{2})^\gamma\,(k+\frac{-k_1+k_2}{2})^\sigma\,(k+\frac{-k_1+k_2}{2})_\mu\nonumber\\
&&\nonumber\\
&&\times\frac{(k-{\displaystyle\frac{k_1+3k_2}{2})_\sigma\,g_{\gamma\nu}+
(k+\frac{-k_1+3k_2}{2})_\gamma\,g_{\nu\sigma}+(-2k+k_1)_\nu\,g_{\gamma\sigma}}}
{\big[\big(k+\frac{k_1+k_2}{2}\big)^2-M^2+i\epsilon\big]\,[\big(k+\frac{-k_1+k_2}{2}\big)^2-M^2+i\epsilon\big]\,
\big[\big(k-\frac{k_1+k_2}{2}\big)^2-M^2+i\epsilon\big]}\,,\nonumber\\
&&\label{eq4.14}
\end{eqnarray}
and
\begin{eqnarray}
{\cal M}_{143}&=&\frac{ie^2gM}{(2\pi)^4}\int
d^4k\,g^\sigma_\mu\,(k+\frac{k_1+k_2}{2})^\gamma\nonumber\\
&&\nonumber\\
&&\times\frac{(k-{\displaystyle\frac{k_1+3k_2}{2})_\sigma\,g_{\gamma\nu}+
(k+\frac{-k_1+3k_2}{2})_\gamma\,g_{\sigma\nu}+(-2k+k_1)_\nu\,g_{\gamma\sigma}}}
{\big[\big(k+\frac{k_1+k_2}{2}\big)^2-M^2+i\epsilon\big]\,[\big(k+\frac{-k_1+k_2}{2}\big)^2-M^2+i\epsilon\big]\,
\big[\big(k-\frac{k_1+k_2}{2}\big)^2-M^2+i\epsilon\big]}\,.\nonumber\\
&&\label{eq4.15}
\end{eqnarray}
Symmetrizing ${\cal M}_{141}$ in the two photons by adding the
analogous part of amplitude~${\cal M}_3$, we obtain
\begin{eqnarray}
{\cal M}_{141}+{\cal
M}_{341}&=&\frac{ie^2gM}{(2\pi)^4}\,\frac{1}{M^2}\int
d^4k\,g^\sigma_\mu\,(k+\frac{k_1+k_2}{2})^\gamma\nonumber\\
&&\nonumber\\
&&\times\frac{(-k-{\displaystyle\frac{k_2}{2})_\mu\,g_{\gamma\nu}+
(2k+k_1+k_2)_\gamma\,g_{\mu\nu}+(-k-\frac{k_1}{2})_\nu\,g_{\mu\gamma}}}
{\big[\big(k+\frac{k_1+k_2}{2}\big)^2-M^2+i\epsilon\big]\,
\big[\big(k-\frac{k_1+k_2}{2}\big)^2-M^2+i\epsilon\big]}\,,\label{eq4.16}
\end{eqnarray}
which is readily seen as the opposite of~${\cal M}_{22}$ in
eq.~\eqref{eq4.9}. Hence,
\begin{equation}
{\cal M}_{141}+{\cal M}_{341}+{\cal M}_{22}=0\,.\label{eq4.17}
\end{equation}

To continue the treatment of the $M^{-2}$ terms, we collect the
remaining terms from~${\cal M}_{113}$, ${\cal M}_{13}$, ${\cal
M}_{122}$, and ${\cal M}_{142}$ as given by eqs.~\eqref{eq3.3},
\eqref{eq4.2}, \eqref{eq4.6}, and~\eqref{eq4.14} respectively:
\begin{eqnarray}
\lefteqn{{\cal M}_{113}+{\cal M}_{13}+{\cal M}_{122}+{\cal
M}_{142}=\frac{-ie^2gM}{(2\pi)^4}\,\frac{1}{M^2}}
\nonumber\\
&&\nonumber\\
&&\times\int d^4k\,\frac{A}
{\big[\big(k+\frac{k_1+k_2}{2}\big)^2-M^2+i\epsilon\big]\,[\big(k+\frac{-k_1+k_2}{2}\big)^2-M^2+i\epsilon\big]\,
\big[\big(k-\frac{k_1+k_2}{2}\big)^2-M^2+i\epsilon\big]}\,,\nonumber\\
&&\label{eq4.18}
\end{eqnarray}
with
\begin{eqnarray}
A&=&{\displaystyle
-\big(k+\frac{k_1+k_2}{2}\big)^2\,\big(k-\frac{k_1+k_2}{2}\big)^2\,g_{\mu\nu}-(k^2-\frac{(k_1\cdot
k_2)}{2})\, (k+\frac{k_2}{2})_\mu\,(k-\frac{k_1}{2})_\nu}\nonumber\\
&&\nonumber\\
&&+{\displaystyle\big(k+\frac{k_1+k_2}{2}\big)^2\,(k-\frac{k_2}{2})_\mu\,(k-\frac{k_1}{2})_\nu
+\big(k-\frac{k_1+k_2}{2}\big)^2\,(k+\frac{k_2}{2})_\mu\,(k+\frac{k_1}{2})_\nu}\nonumber\\
&&\nonumber\\
&&+{\displaystyle(k^2-\frac{(k_1\cdot k_2)}{2})\,\big[\,(k^2-(k\cdot
k_1)+(k\cdot k_2)-\frac{(k_1\cdot
k_2)}{2}\,)g_{\mu\nu}+(k+\frac{k_2}{2})_\mu\,(k-\frac{k_1}{2})_\nu\,]}\nonumber\\
&&\nonumber\\
&&+{\displaystyle(k+\frac{k_2}{2})_\mu\,\big[\,\big(k-\frac{k_1+k_2}{2})^2\,\big(k+\frac{k_1}{2})_\nu-(k+\frac{k_1+k_2}{2}\cdot
k-\frac{k_1+k_2}{2}\big)\,(k-\frac{k_1}{2})_\nu\,\big]}\nonumber\\
&&\nonumber\\
&&+{\displaystyle(k-\frac{k_1}{2})_\nu\,\big[\,\big(k+\frac{k_1+k_2}{2})^2\,\big(k-\frac{k_2}{2})_\mu-(k+\frac{k_1+k_2}{2}\cdot
k-\frac{k_1+k_2}{2}\big)\,(k+\frac{k_2}{2})_\mu\,\big]}\,.\nonumber\\
&&\label{eq4.19}
\end{eqnarray}
Some elementary algebra shows that
\begin{eqnarray}
A&=&4\,(k_1\cdot k_2)\,k_\mu\,k_\nu+2\,k^2\,k_{2\mu}\,k_{1\nu}
-2\,(k_\mu\,k_{1\nu}+k_{2\mu}\,k_\nu)(k\cdot k_1+k_2)\nonumber\\
&&\nonumber\\
&&+g_{\mu\nu}\,[-2\,k^2\,(k_1\cdot k_2)+(k\cdot k_1+k_2)^2\,]\nonumber\\
&&\nonumber\\
&&+[\,k^2-\frac{(k_1\cdot k_2)}{2}\,]\,[-g_{\mu\nu}\,(k\cdot
k_1-k_2)+2\,(k_\mu\,k_{1\nu}-k_{2\mu}\,k_\nu)\,]\,.\label{eq4.20}
\end{eqnarray}
We want to rewrite the last line in this expression for $A$
as follows:
\begin{eqnarray}
\lefteqn{[\,k^2-\frac{(k_1\cdot k_2)}{2}\,]\,[-g_{\mu\nu}\,(k\cdot
k_1-k_2)+2\,(k_\mu\,k_{1\nu}-k_{2\mu}\,k_\nu)\,]}\nonumber\\
&&\nonumber\\
&\hspace*{2cm}=&[\,k^2-(k\cdot k_1-k_2)-\frac{(k_1\cdot k_2)}{2}-M^2 +(k\cdot
k_1-k_2)+M^2\,]\nonumber\\
&&\nonumber\\
&&\times\,[-g_{\mu\nu}\,(k\cdot
k_1-k_2)+2\,(k_\mu\,k_{1\nu}-k_{2\mu}\,k_\nu)\,]\,,\label{eq4.21}
\end{eqnarray}
the reason being that the first four terms in the first bracket
cancel the middle denominator in~\eqref{eq4.17}, which makes the
denominator an even function of~$k$. As the second bracket is odd
in~$k$, we can drop that term entirely. In the process, we have
introduced a term in the numerator proportional to~$M^2$, which will
have to be treated in the next subsection~\ref{sec3.4}, together
with the remaining terms. Thus,
\begin{equation}
{\cal M}_{113}+{\cal M}_{13}+{\cal M}_{122}+{\cal M}_{142}={\cal
M}_{1131}+{\cal M}_{1132}\,.\label{eq4.22}
\end{equation}
In eq.~\eqref{eq4.22}, we have
\begin{eqnarray}
\lefteqn{{\cal M}_{1131}=\frac{-ie^2gM}{(2\pi)^4}\,\frac{1}{M^2}}
\nonumber\\
&&\nonumber\\
&&\times\int d^4k\,\frac{A'}
{\big[\big(k+\frac{k_1+k_2}{2}\big)^2-M^2+i\epsilon\big]\,[\big(k+\frac{-k_1+k_2}{2}\big)^2-M^2+i\epsilon\big]\,
\big[\big(k-\frac{k_1+k_2}{2}\big)^2-M^2+i\epsilon\big]}\,,\nonumber\\
&&\label{eq4.23}
\end{eqnarray}
with
\begin{eqnarray}
A'&=&4\,(k_1\cdot k_2)\,k_\mu\,k_\nu+2\,k^2\,k_{2\mu}\,k_{1\nu}
-4\,k_\mu\,k_{1\nu}\,(k\cdot k_2)-4\,k_{2\mu}\,k_\nu\,(k\cdot k_1)\nonumber\\
&&\nonumber\\
&&+g_{\mu\nu}\,[-2\,k^2\,(k_1\cdot k_2)+4\,(k\cdot k_1)\,(k\cdot
k_2)\,]\,,\label{eq4.24}
\end{eqnarray}
and
\begin{eqnarray}
\lefteqn{{\cal M}_{1132}=\frac{-ie^2gM}{(2\pi)^4}}\nonumber\\
&&\nonumber\\
&&\times\int d^4k\,\frac{-g_{\mu\nu}\,(k\cdot
k_1-k_2)+2\,(k_\mu\,k_{1\nu}-k_{2\mu}\,k_\nu)}
{\big[\big(k+\frac{k_1+k_2}{2}\big)^2-M^2+i\epsilon\big]\,[\big(k+\frac{-k_1+k_2}{2}\big)^2-M^2+i\epsilon\big]\,
\big[\big(k-\frac{k_1+k_2}{2}\big)^2-M^2+i\epsilon\big]}\,,\nonumber\\
&&\label{eq4.25}
\end{eqnarray}

We proceed to show that the expression~${\cal M}_{1131}$ vanishes.
To this end, note that the integral in~\eqref{eq4.23} is only
logarithmically divergent. It follows that a shift in the
integration variable is allowed and that it does not produce any
surface term. Combining the three factors in the denominator with
the Feynman variables $\alpha_1$, $\alpha_2$, and $\alpha_3$, we
obtain the denominator
\begin{eqnarray}
D&=&k^2+\alpha_1\,(k\cdot k_1+k_2)+\alpha_3\,(k\cdot
-k_1+k_2)-\alpha_2\,(k\cdot k_1+k_2)\nonumber\\
&&\nonumber\\
&&+(\alpha_1-\alpha_3+\alpha_2)\,\frac{(k_1\cdot k_2)}{2}
-M^2+i\epsilon\nonumber\\
&&\nonumber\\
&=&k^2-(1-2\,\alpha_1)\,(k\cdot k_1)+(1-2\,\alpha_2)\,(k\cdot
k_2)+(1-2\,\alpha_3)\,\frac{(k_1\cdot
k_2)}{2}-M^2+i\epsilon\,,\nonumber\\
&&\label{eq4.26}
\end{eqnarray}
because of the $\delta$-function
$\delta(1-\alpha_1-\alpha_2-\alpha_3)$ in the Feynman combination of
denominators. With the shift
\begin{equation}
k=\ell+\frac{1}{2}(1-2\,\alpha_1)\,k_1-\frac{1}{2}(1-2\,\alpha_2)\,k_2
,,\label{eq4.27}
\end{equation}
the denominator becomes
\begin{equation}
D=\ell^2-M^2+2\,\alpha_1\,\alpha_2\,(k_1\cdot
 k_2)+i\epsilon\,.\label{eq4.28}
\end{equation}
It is a simple matter to perform the shift~\eqref{eq4.27} in the
numerator~$A'$ in~\eqref{eq4.24}. Dropping the terms odd in~$\ell$
and using the relations~\eqref{eq2.4}, one finds that effectively
$A'=0$, meaning that our amplitude does not contain terms
in~$M^{-2}$. This result is closely related to the fact that our
amplitude obeys the decoupling theorem.

\subsection{The result}\label{sec3.4}
The last set of terms to be treated are those without negative
powers of~$M$. First, we list the term that derives from the
amplitude~${\cal M}_1$ as given by~\eqref{eq2.1}:
\begin{eqnarray}
{\cal M}_{15}&=&\frac{-ie^2gM}{(2\pi)^4}\,\int
d^4k\,g^{\rho\sigma}\,g^{\beta\gamma}\nonumber\\
&&\nonumber\\
&&\times[\,(k+\frac{3k_1+k_2}{2})_\rho\, g_{\beta\mu}+(k+\frac{-3k_1+k_2}{2})_\beta\, g_{\mu\rho}+(-2k-k_2)_\mu\, g_{\rho\beta}\,]\nonumber\\
&&\nonumber\\
&&\times\frac{(k-{\displaystyle\frac{k_1+3k_2}{2})_\sigma\,g_{\gamma\nu}+(k+\frac{-k_1+3k_2}{2})_\gamma\,
g_{\nu\sigma}+(-2k+k_1)_\nu\,g_{\sigma\gamma}}}
{\big[\big(k+\frac{k_1+k_2}{2}\big)^2-M^2+i\epsilon\big]\,\big[\big(k+\frac{-k_1+k_2}{2}\big)^2-M^2+i\epsilon\big]\,
\big[\big(k-\frac{k_1+k_2}{2}\big)^2-M^2+i\epsilon\big]}\,,\nonumber\\
&&\nonumber\\
&&\nonumber\\
&=&\frac{-ie^2gM}{(2\pi)^4}\,\int
d^4k\,\Big\{\,g_{\mu\nu}\,[\,2\,k^2-(k\cdot k_1)+(k\cdot k_2)-5\,(k_1\cdot k_2)\,]\nonumber\\
&&\nonumber\\
&&\frac{\hspace*{4.5cm}+{\displaystyle
2\,k_\mu\,k_\nu+\frac{9}{2}\,k_{2\mu}\,k_{1\nu}+8\,(k+\frac{k_2}{2})_\mu\,(k-\frac{k_1}{2})_\nu\,\Big\}}}
{\big[\big(k+\frac{k_1+k_2}{2}\big)^2-M^2+i\epsilon\big]\,\big[\big(k+\frac{-k_1+k_2}{2}\big)^2-M^2+i\epsilon\big]\,
\big[\big(k-\frac{k_1+k_2}{2}\big)^2-M^2+i\epsilon\big]}\,,\nonumber\\
&&\label{eq5.1}
\end{eqnarray}
and from~${\cal M}_2$ as given by~\eqref{eq2.2}
\begin{eqnarray}
{\cal M}_{24}&=&\frac{ie^2gM}{(2\pi)^4}\,\int d^4k\,
\frac{g^{\beta\gamma}\,[\,2\,g_{\mu\nu}\,g_{\beta\gamma}-g_{\mu\beta}\,g_{\nu\gamma}-g_{\mu\gamma}\,
g_{\nu\beta}\,]}
{\big[\big(k+\frac{k_1+k_2}{2}\big)^2-M^2+i\epsilon\big]\,
\big[\big(k-\frac{k_1+k_2}{2}\big)^2-M^2+i\epsilon\big]}
\nonumber\\
&&\nonumber\\
&&\nonumber\\
&=&\frac{ie^2gM}{(2\pi)^4}\,\int d^4k\, \frac{6\,g_{\mu\nu}}
{\big[\big(k+\frac{k_1+k_2}{2}\big)^2-M^2+i\epsilon\big]\,
\big[\big(k-\frac{k_1+k_2}{2}\big)^2-M^2+i\epsilon\big]}
\,.\label{eq5.2}
\end{eqnarray}
The remaining contributions are ${\cal M}_{123}$, ${\cal M}_{143}$,
and ${\cal M}_{1132}$ as given by eqs.~\eqref{eq4.7},
\eqref{eq4.15}, and~\eqref{eq4.25} respectively. Adding these
contributions gives
\begin{eqnarray}
\lefteqn{{\cal M}_{15}+\frac{1}{2}{\cal M}_{24}+{\cal M}_{123}+{\cal
M}_{143}+{\cal M}_{1132}=\frac{-ie^2gM}{(2\pi)^4}\,\int
d^4k}\nonumber\\
&&\nonumber\\
&&\Big\{\,g_{\mu\nu}\,[\,-3\,k^2+3\,(k\cdot k_1)-3\,(k\cdot k_2)-\frac{9}{2}\,(k_1\cdot k_2)+3\,M^2\,]\nonumber\\
&&\nonumber\\
&&\frac{\hspace*{4.8cm}{\displaystyle
+12\,k_\mu\,k_\nu+3\,k_{2\mu}\,k_{1\nu}-6\,k_\mu
k_{1\nu}+6\,k_{2\mu}\,k_\nu\,\Big\}}}
{\big[\big(k+\frac{k_1+k_2}{2}\big)^2-M^2+i\epsilon\big]\,\big[\big(k+\frac{-k_1+k_2}{2}\big)^2-M^2+i\epsilon\big]\,
\big[\big(k-\frac{k_1+k_2}{2}\big)^2-M^2+i\epsilon\big]}\,,\nonumber\\
&&\label{eq5.3}
\end{eqnarray}

We perform the same shift of integration variable as in
eq.~\eqref{eq4.24}, we drop the odd terms in~$\ell$, and we
symmetrize
\begin{equation}
\ell_\mu\,\ell_\nu\rightarrow
\frac{1}{4}\,\ell^2\,g_{\mu\nu}\,,\label{eq5.4}
\end{equation}
leading to
\begin{eqnarray}
\lefteqn{{\cal M}_{15}+\frac{1}{2}{\cal M}_{24}+{\cal M}_{123}+{\cal
M}_{143}+{\cal M}_{1132}=\frac{-2ie^2gM}{(2\pi)^4}\,\int
d^4\ell}\nonumber\\
&&\nonumber\\
&&\int_0^1d\alpha_1\int_0^{1-\alpha_1}d\alpha_2\,\,\frac{\,g_{\mu\nu}\,[\,(k_1\cdot
k_2)\,(-6+6\,\alpha_1\alpha_2)+3\,M^2\,]+3\,(2-4\,\alpha_1\alpha_2)\,k_{2\mu}\,k_{1\nu}}
{[\,\ell^2-M^2+2\alpha_1\alpha_2\,(k_1\cdot
k_2)+i\epsilon\,]^3}\,.\nonumber\\
&&\label{eq5.5}
\end{eqnarray}

Performing the integration over~$d^4\ell$, and adding the
contribution from~${\cal M}_3$ (and the other half of ${\cal M}_2$),
which merely gives a factor of 2 because the result~\eqref{eq5.5} is
already $1\leftrightarrow2$ symmetric, we obtain the total
amplitude~${\cal M}$
\begin{eqnarray}
\lefteqn{{\cal M}={\cal M}_1+{\cal M}_2+{\cal M}_3=\frac{-e^2gM}{8\pi^2}}\nonumber\\
&&\nonumber\\
&&\times\int_0^1d\alpha_1\int_0^{1-\alpha_1}d\alpha_2\,\,\frac{g_{\mu\nu}\,[\,(k_1\cdot
k_2)\,(-6+6\,\alpha_1\alpha_2)+3\,M^2\,]+3\,(2-4\,\alpha_1\alpha_2)\,k_{2\mu}\,k_{1\nu}}
{M^2-2\,\alpha_1\alpha_2\,(k_1\cdot
k_2)-i\epsilon}\,.\nonumber\\
&&\label{eq5.6}
\end{eqnarray}
Following Dyson's prescription~\cite{R16}, we perform a subtraction
of the amplitude for~\mbox{$k_1=k_2=0$}, to obtain the finite and
gauge invariant result
\begin{equation}
{\cal M}=\frac{-e^2gM}{8\pi^2}\int_0^1d\alpha_1\int_0^{1-\alpha_1}d\alpha_2\,\,
\frac{6\,(1-2\,\alpha_1\alpha_2)\,[\,k_{2\mu}\,k_{1\nu}-g_{\mu\nu}\,(k_1\cdot
k_2)\,]}{M^2-\alpha_1\alpha_2\,M_H^2-i\epsilon}\,.\label{eq5.7}
\end{equation}

The integral over the Feynman parameters can be expressed in terms
of elementary functions. Using
\begin{equation}
\tau=\frac{M_H^2}{4M^2}\,,\label{eq5.8}
\end{equation}
we have
\begin{equation}
{\cal M}=-\frac{3e^2g}{8\pi^2M}\,[\,k_{2\mu}\,k_{1\nu}-g_{\mu\nu}\,(k_1\cdot
k_2)\,]\,[\,\tau^{-1}+(2\,\tau^{-1}-\tau^{-2})\,f(\tau)\,]\,,\label{eq5.9}
\end{equation}
with
\begin{equation}
f(\tau)=\left\{\begin{array}{lcc}
\arcsin^2(\sqrt{\tau})&\mbox{for}&\tau\leq 1\,,\\[5mm]
-{\displaystyle\frac{1}{4}\,\left[\ln\frac{1+\sqrt{1-\tau^{-1}}}{1-\sqrt{1-\tau^{-1}}}-i\pi\right]^2}
&\mbox{for}&\tau>1\,.\end{array}\right.\label{eq5.10}
\end{equation}
Clearly, for large Higgs masses ($\tau\rightarrow\infty$), we have
from eq.~\eqref{eq5.9} that ${\cal M}\rightarrow0$, \mbox{i.e.}, we
have decoupling of the $W$ contribution~\cite{R7}.

\section{Discussions and conclusions}\label{sec4}
Several points concerning our result eq.~\eqref{eq5.9} deserve
further discussion.
\renewcommand{\labelenumi}{(\alph{enumi})}
\begin{enumerate}
\item As announced in the Introduction, our result~\eqref{eq5.9}
differs from the previous one~\cite{R3}, which reads
\begin{equation}
{\cal M}(\xi=1)=-\frac{e^2g}{8\pi^2M}\,[\,k_{2\mu}\,k_{1\nu}-g_{\mu\nu}\,(k_1\cdot
k_2)\,]\,[\,2+3\,\tau^{-1}+3\,(2\,\tau^{-1}-\tau^{-2})\,f(\tau)\,]\,.\label{eq5.11}
\end{equation}
The first term in the second bracket of~\eqref{eq5.11} is not
present in our result~\eqref{eq5.9}. It is precisely the one which
violates the decoupling theorem because it does not vanish
for~$\tau\rightarrow\infty$.

The observation that two honest calculations for the same process in
quantum field theory can lead to two different answers is both
disturbing and intriguing. It is disturbing because the question
naturally arises: which is the right answer from the physics point
of view? One rather compelling argument in favor of our answer is
the fact that our amplitude does satisfy the decoupling theorem for
large Higgs masses.

In our opinion, the present calculation is reliable because it is
straightforward: it is a calculation that is convergent and,
therefore, it does not appeal to artifacts such as regularization.

\item In the pioneering paper of Ellis, Gaillard, and Nanopoulos~\cite{R2} on
the decay process~\eqref{eq1.1} through one~$W$ loop, the
assumption~(c) of Sec.~\ref{sec1} was used: at that time,
thirty-five years ago, it was believed that the Higgs particle had a
small mass. In this limit of small Higgs mass, the present result
given by eq.~\eqref{eq5.9} is smaller by a factor of~5/7 compared
with that of Ref.~\cite{R2}. In the opposite limit of a large Higgs
mass compared with the mass of the~$W$, the present result, which
satisfies the decoupling theorem~\cite{R7}, is in absolute value
much smaller than the previous one~\cite{R2,R3}.

\item For a Higgs mass $M_H=115$~GeV$/c^2$, the quantity $\tau=0.511\,$
[see eq.~\eqref{eq5.8}]. A comparison of the two
expressions~\eqref{eq5.9} and~\eqref{eq5.11} then shows that the
amplitude~${\cal M}$ is~24.9\% smaller in absolute value than
the~${\cal M}(\xi=1)$ amplitude. If the top loop is also taken into
account~\cite{R18}, the decay width for $H\rightarrow\gamma\gamma$
is reduced by~54.2\%.

\item The origin of the extra term~$2$ in~\eqref{eq5.11} can be
traced to the use of dimensional regularization. The previous
calculations~\cite{R3} were indeed performed in the~$R_1$ gauge
with an implementation of dimensional regularization.

This regularization scheme requires that the algebra be performed
in~$n$ dimensions. Hence, the symmetrization~\eqref{eq5.4}
is to be replaced by
\begin{equation}
\ell_\mu\,\ell_\nu\rightarrow
\frac{1}{n}\,\ell^2\,g_{\mu\nu}\label{eq5.12}
\end{equation}
and one must take
\begin{equation}
g_\mu^\mu = n\,. \label{eq5.12a}
\end{equation}

It is readily verified that the additional $n$-dependence from eq.~\eqref{eq5.12a}
does not change the result~\eqref{eq5.11}. Therefore, the difference
between the two results~\eqref{eq5.9}
and~\eqref{eq5.11} stems from the behavior of the following
integral:
\begin{equation}
I_{\mu\nu}(n)=\int
d^n\ell\,\frac{\ell^2\,g_{\mu\nu}-4\,\ell_\mu\,\ell_\nu}{[\,\ell^2-M^2+i\epsilon\,]^3}\,.\label{eq5.13}
\end{equation}
By symmetric integration~\eqref{eq5.4}, this integral is
\begin{equation}
I_{\mu\nu}(4)=0\,;\label{eq5.14}
\end{equation}
on the other hand, a direct evaluation with~\eqref{eq5.12} yields
\begin{equation}
I_{\mu\nu}(n)\simeq-\dfrac{i\pi^2}{2}\,g_{\mu\nu}\,,\label{eq5.15}
\end{equation}
when~$n$ is close to but less than~4. Thus, the
integral~$I_{\mu\nu}(n)$ is discontinuous at~$n=4$.

Is such a behavior ``pathological''? In view of the simple nature of
the integral~\eqref{eq5.13} and the fact that this integral is not
defined for~$n>4$, the answer must be ``no''. This raises the
question how often dimensional regularization can lead to wrong
answers. For the present case, the previous result is suspicious
because of its failure to satisfy the decoupling theorem; in other
cases, there may be no such guidance to suggest the necessity of
repeating the calculation with the space-time dimension kept at
four.

\end{enumerate}

\medskip

\section*{Acknowledgments}

\noindent One of us (T.T.W.) is greatly indebted to the CERN Theory
Group for their hospitality.

\newpage

\end{document}